\documentclass[superscriptaddress,secnumarabic,
amssymb,amsmath,nobibnotes,aps,prd,showkeys,showpacs,nofootinbib]{revtex4}%
\usepackage{graphicx}
\usepackage{epsf}
\usepackage{bm}
\usepackage{amsmath}
\usepackage{amsfonts}
\usepackage{amssymb}
\usepackage{hyperref}
\usepackage{epstopdf}
\usepackage{natbib}%
\usepackage{epsf,graphics,graphicx}
\usepackage{amsmath}
\usepackage{amssymb,latexsym,mathrsfs, bm}
\usepackage{color}
\providecommand{\U}[1]{\protect\rule{.1in}{.1in}}
\newcommand{\be}{\begin{equation}}
\newcommand{\ee}{\end{equation}}

\newcommand{\mincir}{\raise
-3.truept\hbox{\rlap{\hbox{$\sim$}}\raise4.truept\hbox{$<$}\ }}
\newcommand{\magcir}{\raise
-3.truept\hbox{\rlap{\hbox{$\sim$}}\raise4.truept\hbox{$>$}\ }}

\begin{document}
\title{Gravitationally influenced particle creation models and late-time cosmic acceleration}
\author{Supriya Pan}
\email{span@research.jdvu.ac.in}
\affiliation{Department of Physical Sciences, Indian Institute of Science Education and Research Kolkata, Mohanpur$-$741246, West Bengal, India}
\author{Barun Kumar Pal}
\email{terminatorbarun@gmail.com}
\affiliation{Department of Mathematics, Netajinagar College for Women, Kolkata- 700092, West Bengal, India}
\author{Souvik Pramanik}
\email{souvik.in@gmail.com}
\affiliation{Physics and Applied Mathematics Unit, Indian Statistical Institute, 203 B.T. Road, Kolkata-700108, West Bengal, India}
\begin{abstract}
In this work we focus on the gravitationally influenced adiabatic particle creation process, a mechanism that does not need any dark energy or modified gravity models to explain the current accelerating phase of the universe. Introducing some particle creation models that generalize some previous models in the literature, we constrain the cosmological scenarios using the latest compilation of the Type Ia Supernovae data only, the first indicator of the accelerating universe. Aside from the observational constraints on the models, we examine the models using two model independent diagnoses, namely the cosmography and $Om$. Further, we establish the general conditions to test the thermodynamic viabilities of any particle creation model. Our analysis shows that at late-time, the models have close resemblance to that of the $\Lambda$CDM cosmology, and the models always satisfy the generalized second law of thermodynamics under certain conditions. 
\end{abstract}

\maketitle
\section{Introduction}
\label{Intro}
Almost two decades have elapsed since the exploitation of the fact that our universe is accelerating by measuring the luminosity distance of the type Ia Supernovae \cite{Riess1}. Since then many observational evidences from different cosmic sources \cite{Perlmutter1, Komatsu1, Sanchez1} have put the opine ``universe is currently accelerating'' into a firm observational footing. This acceleration can be described as an effect of some unknown component(s), named as dark energy which is completely  unknown by its character, origin and it needs large negative pressure to accelerate. Among other dark energy candidates, $\Lambda$CDM has been found to depict the  contemporary observations at the best. However, this cosmological model suffers from two major drawbacks $-$ cosmological constant problem \cite{Weinberg1}  and the cosmic coincidence problem \cite{Steinhardt2003}. These drawbacks forced the grudging cosmologists to tour beyond  $\Lambda$CDM model. As a consequence, 
the dynamical  nature of dark energy was proposed.
Being the simplest such candidate,
the scalar field(s) dark energy models came into existence to explain this late-time cosmic acceleration. As of now there are plenty of scalar field driven models in the literature, such as, quintessence, k-essence, tachyon, phantom, quintom etc (see for instance \cite{CST}). These scalar field models are  attractive due to their ability to produce cosmic acceleration in agreement with current observational data, however, the cosmic coincidence problem was soon attached with such models. The modified gravity models are also considered for a possible explanation to the current acceleration of our universe \cite{Cap,Carroll04,NO, Nojiri:2010wj,Capozziello:2011et, Nojiri:2017ncd}. For a detailed description of the thermodynamic properties of the modified gravity models, we refer \cite{Bamba:2016aoo} and the references cited therein. Apart from the dark energy and modified gravity theories, there are some other explanations to the current acceleration. In  Ref.\cite{Kleidis:2011ga}   it has been argued that the inclusion of collisional dark matter may result in accelerating universe. Moreover, peculiar velocities of the relativistic observers could be a possible explanation to the present accelerated expansion of the universe, see \cite{Tsagas:2011wq} where it has been discussed that due to such peculiar velocities, the observers in a particular galaxy could experience a local accelerating expansion while the global behaviour of the universe may be decelerating. The motivation of the current work is somewhat different from the above theories that we going to describe in the next section.  

Recently, a great attention has been paid on the cosmology of gravitationally induced `adiabatic' particle production since they successfully explain the current accelerated expansion \cite{SSL09,LJO10,LGPB14,RSW14,FPP14,CPS2014,NP15,LSC2016, NP2016, HP2015, PHPS16}. This field is very appealing for several important observations carried our in the last couple of years. The particle productions of light non-minimally
coupled scalar fields due to the change in the spacetime geometry can lead to an early accelerating universe \cite{Sanhi-Habib}. In addition, quantum corrections could also lead to particle productions in an early universe which may result in an equation of state $w< -1$ \cite{Onemli2002, Onemli2004, Onemli2007}, that means a realization of the phantom universe without invoking any scalar field is not necessary. The connection between early inflationary era and the present acceleration of the universe is also pointed out in Ref. \cite{RCN2016}. In this work we only concentrate on the late time acceleration of our universe, and we aim to provide a concise description of such a theory in short. We start with recalling the microscopic formulation of particle productions by the gravitational field in 1939 by Schr\"{o}dinger \cite{Schrodinger}. After a long period, Parker and collaborators \cite{Parker} re-investigated this microscopic formulation of particle productions built on the Bogoliubov mode-mixing technique in
the background of Quantum Field Theory (QFT) in a curved space-time \cite{Birrell}. After such microscopic investigations of the particle productions, its macroscopic description was studied by Prigogine et al \cite{Prigogine1} based on the non-equilibrium thermodynamics where the universe was assumed to be an open thermodynamical system, and they were successful to connect the particle productions into the Einstein's field equations in a consonant manner. However, this approach was further investigated by a covariant formulation \cite{Calvao1} and applied to cosmology where the back reaction term is naturally absorbed into the Einstein's field equations leading to a negative pressure which is responsible for the current cosmic acceleration. However, in connection with this particle productions at the expense of gravitational field of this expanding universe, we recall that long back ago, Zeldovich \cite{Zeldovich1970} introduced some bulk viscosity mechanism which is responsible for particle productions.
However, later on, Lima and Germano \cite{LG92a} (also see \cite{LG92b}) showed that although both the processes, such as, bulk viscosity mechanism by Zeldovich \cite{Zeldovich1970} and gravitational particle productions produces same dynamics of the universe but in principle, they are completely different from a thermodynamical point of view. Since we are describing the particle productions, we would also like to note an analogy which exists between the models driven by particle productions and the models of Steady State Cosmology developed in \cite{HN}. But, both the approaches are again different form their construction since the Steady State Cosmological models are inspired by adding extra terms into the Einstein-Hilbert action interpreting the so-called C-filed, and the  creation phenomenon is comprehended through a process of exchanging the energy and momentum between matter itself and the C-field.

In the gravitationally induced particle creation mechanism, the usual balance equation $N_{;\mu}^{\mu}= 0$, is modified  as \cite{Prigogine1}
	\begin{equation}
	N_{;\mu}^{\mu}\equiv\mathit{n}_{,\mu} u^{\mu}+\Theta\mathit{n}=\mathit{n}
	\Gamma~\Longleftrightarrow~N_{,\mu}u^{\mu}=\Gamma N, \label{balance-eqn}
	\end{equation}
	where $N^{\mu} =\mathit{n}u^{\mu}$ is the particle flow vector; $u^{\mu}$ is the usual four velocity of the created particles;
	$\Gamma$ is the rate of change of the particle number in a physical volume $V$ containing $N$ number of particles,  $\mathit{n}=N/V$, is the particle number density, $\Theta=u_{;\mu}^{\mu}$, denotes the fluid expansion. Hence, due to the modifications in the balance equation (\ref{balance-eqn}), the field equations $G_{\mu\nu}= 8 \pi G T_{\mu\nu}$ will be modified accordingly, and the modified field equations will describe the state of the universe in presence of the creation of particles by the gravitational field.

Now, the key of the dynamics is to find an exact functional form of the particle creation rate $\Gamma$ which could mimic the current observation. But, the possibility of having such a correct functional form of $\Gamma$ can never be realized until a proper description of QFT in the Friedmann-Lema{\^\i}tre- Robertson-Walker (FLRW) universe is available. Therefore, in general one starts with some choices for the production  rate and fit the associated cosmological parameters with the current astronomical data (see for instance \cite{SSL09,LJO10,LGPB14,RSW14,FPP14,CPS2014,NP15,LSC2016, NP2016}).  Here we have considered three most general phenomenological rate $\Gamma$ and investigate the evolution equation by Union 2.1 data to see how well they depict present astronomical data. We found that all the models predict a smooth transition from decelerating phase to the current accelerating phase at around $z \sim 1$. Then we have employed two model independent tests, namely, the cosmography and the $Om$ diagnostic
so that, we can measure the deviation of the models from the $\Lambda$CDM, as this is the best description for our  universe till date. \newline

 The paper is organized as follows: In section \ref{field-equations}, presenting the field equations, we have introduced our three particle production models and analyzed them by Union 2.1 data. In section \ref{Sec-CP}, we have  discussed the model independent tests, cosmography and $Om$ respectively. Then we have presented the thermodynamic analysis of the models in section \ref{sec:thermodynamics}. Finally, in the last section \ref{discussion}, we have discussed  the outcomes of our work. We note that throughout the paper, we have used particle productions and matter creation synonymously.

\section{Field equations in FLRW universe}
\label{field-equations}
In agreement with cosmological inflation and the cosmic microwave background radiation, the geometry of our universe  is very well described by the FLRW line element, which for zero spatial curvature is given by

\begin{equation}
ds^2= -dt^2+a^2(t) (dx^2+ dy^2+dz^2),\label{flrw}
\end{equation}
where $a(t)$ is the scale factor of the universe. In this background, the non-trivial Einstein's field equations for a perfect fluid endowed with gravitationally induced matter creation model are given by
\begin{eqnarray}
H^2&=& \frac{8 \pi G}{3}\,\, \rho,\label{EFE1}\\
\dot{H}+ H^2&=& -\frac{4 \pi G}{3}\, (\rho+ 3p+3p_c),\label{EFE2}
\end{eqnarray}
where $\rho$, $p$ are respectively the energy density and the thermodynamic pressure of the perfect fluid, $p_c$ denotes the  creation pressure due to the gravitationally induced particle productions, and the over dot represents the cosmic time differentiation. Now, under `adiabatic' condition, this $p_c$ can be written as \cite{Prigogine1,Calvao1,HP2015}
\begin{eqnarray}
p_c&=& -\frac{\Gamma}{3H} (p+\rho),
\end{eqnarray}
where $H= \dot{a}/a$ is the usual Hubble rate, $\Gamma$ is the rate of matter creation from the gravitational field. In principle, $\Gamma> 0$ represents the matter creation, $\Gamma< 0$ is for matter annihilation, and $\Gamma= 0$ is the case when there is no matter creation. But, the validity of the generalized thermodynamics in such a scenario induces $\Gamma > 0$. In general, the exact form of $\Gamma$ is unknown, but it should be determined in the context of quantum processes in curved
space time and by taking into account the back
reaction effects. Note that, for an expanding universe, the creation pressure $p_c$ is  negative.

In what follows, we consider that the perfect fluid satisfying the barotropic equation of state
\begin{eqnarray}\label{barotropic}
p&=& w \rho,
\end{eqnarray}
where $w \geq 0$ is the equation of state parameter of the perfect fluid. Thus, $w= 1/3$ represents the radiation dominated era, whereas $w= 0$ stands for non-relativistic matter. Now, combining the Einstein's field equations (\ref{EFE1}), (\ref{EFE2}) and the barotropic equation of state in Eq. (\ref{barotropic}), we find
\begin{eqnarray}\label{balance}
\dot{\rho}+ 3 H (1+w) \left(1-\frac{\Gamma}{3H}\right) \rho&=& 0,
\end{eqnarray}
which is nothing but the energy conservation equation and could also
be obtained directly from the Bianchi's identity $T^{\mu \nu}_{;\nu}= 0$ (where $T^{\mu \nu}= (\rho+ p+ p_c)u^{\mu}u^{\nu}+(p+p_c)g^{\mu\nu}$). However, for $\Gamma \ll 3H$, we recover the original energy conservation equation $\dot{\rho}+ 3 H (1+w) \rho= 0$, showing no matter creation effect. Now, combining (\ref{EFE1}), (\ref{barotropic}) and (\ref{balance}), we have the following evolution equation
\begin{equation}
\frac{dH}{dt}+ \frac{3}{2} (1+w) H^2 \left(1-\frac{\Gamma}{\Theta}\right)=0,\label{eqn10.1}
\end{equation}
and the deceleration parameter which measures the decelerating/accelerating phase of the universe takes the form
\begin{eqnarray}
q&=& -\left(1+\frac{\dot{H}}{H^2}\right)=-1+ \frac{3}{2}(1+w) \left(1- \frac{\Gamma}{\Theta}\right).\label{eqn11}
\end{eqnarray}
Now, the dynamics of the universe can only be surveyed after the particle creation rate $\Gamma$ is known. The particle production rate is related to the nature of the produced particles under this adiabatic mechanism, and unfortunately, the nature of the produced particles is unknown to us as the associated QFT is yet to be developed which is an essential tool to determine this $\Gamma$. But, as particle production mechanism has become one of the possible alternatives to explain the current expanding accelerating phase of the universe, so we start with some phenomenological but general choices for $\Gamma$. It has been shown that, $\Gamma \propto H^2$ \cite{AL96, Gunzig98} gives the inflationary solution, $\Gamma \propto H$ \cite{PC2015} can explain the intermediate decelerating era, and even $\Gamma= $ constant, can explain the evolution of the universe from big bang to the present accelerating stage \cite{HP2015} \footnote{In this context we mention that although bulk viscosity and the particle creation process are not equivalent from themrdoynamical ground \cite{LG92a}, but both of them could provide with similar results, see \cite{JP2015}. }. Further, a linear combination of the above rates \cite{PHPS16} predicts two accelerating
phases of the universe, one at early phase of its evolution and the second one stands for the present day acceleration. Furthermore, the same linear combination of the above rates claims the decelerating nature of our universe in future \cite{CPS2014}, as also predicted in some other contexts \cite{Carvalho2006, GL2011}. Therefore, we begin with some phenomenological but general choices for $\Gamma$ of the form $\Gamma= 3 \beta H f (z)$ where $f(z)$ is any arbitrary function of the redshift $z$, and $\beta$ is any free non-negative parameter describing the rate of the creation. In order to investigate their viabilities on the onset of late-time accelerating phase of the universe, we start with the following three choices for $\Gamma$:
\begin{eqnarray}
\mbox{Model I:}~~~~~~\Gamma&=& \frac{3 \beta H}{1+z}\tanh\left({\alpha \over 1+z}\right),\label{M1}\\
\mbox{Model II:}~~~~~~\Gamma&=& \frac{3 \beta H}{1+z} \tanh(\alpha (1+z)),\label{M2}\\
\mbox{Model III:}~~~~~~\Gamma&=& 3 \beta H \Bigl(\frac{1}{1+z}\Bigr).\label{M3}
\end{eqnarray}
These models contain only two parameters $\alpha$, and $\beta$. We note that,
	at a particular redshift, the creation rate is solely dependent on $\beta$, although we have another parameter, $\alpha$, but this does not play any significant role to increase or decrease the particle creation rate due to very slowly varying nature of the function $\tanh (x)$ ($x \in \mathbb{R}$). Further, we note that, it is not possible for $\beta$ to be infinitely large, because in that case, the production of particles will be very large and the evolution of cold dark matter particles could exceed the standard evolution law $a^{-3}$, which will contradict the present observations. Hence, the only restriction on the particle creation models is that, the free parameter $\beta$ should take its value in such as way so that $\Gamma/3H \leq 1$. In the following subsections we discuss the pros and cons of the above models.

Solving the evolution equation (\ref{eqn10.1}) for (\ref{M1}), 
(\ref{M2})  and (\ref{M3}), we find

\begin{eqnarray}
H= H_0 (1+z)^{\frac{3}{2} (1+w)} \exp\left(-\frac{3}{2} (1+w) \beta \int_0^z \frac{\tanh\left(\frac{\alpha}{1+z}\right)}{(1+z)^2}\right),
\end{eqnarray}
for model I while 

\begin{eqnarray}
H= H_0 (1+z)^{\frac{3}{2} (1+w)} \exp\left(-\frac{3}{2} (1+w) \beta \int_0^z \frac{\tanh(\alpha (1+z))}{(1+z)^2}\right),
\end{eqnarray}
for model II and finally the Hubble rate for model III becomes 	

\begin{eqnarray}
H= H_0 (1+z)^{\frac{3}{2}(1+w)} \exp\left(-\frac{3\beta}{2}(1+w) \frac{z}{1+z}\right).
\end{eqnarray}

\section{Union 2.1 data and the results}
\label{sec-data-results}

Supernove Type Ia are the first astronomical data thatthat signaled the accelerating expansion of the universe and hence the existence of some dark energy fluid in the universe sector.  Now, 
corresponding to each Type Ia
Supernova (SN Ia) we observe its redshift $z$ and its apparent magnitude $m_{obs}$, which in terms of the distance modulus $\mu_{obs}$ can be calculated as \cite{Suzuki2012} 
 \begin{equation}
  \mu_{obs}=m_{obs}(z)-M+\bar{\alpha} x_1-\bar{\beta} c+\bar{\delta} P,
  \label{muobs} 
 \end{equation}
where $M$ is the absolute magnitude of SN Ia;  
$c$ stands for the Sn Ia color; $x_1$ is the corrections 
connected with deviations from mean values of the lightcurve 
shape, $P$ stands for a mass of a host galaxy. 
The parameters $M$, $\bar{\alpha}$, $\bar{\beta}$ and
$\bar{\delta}$ are  described by the nuisance parameters \cite{Suzuki2012}. 
Here to analyze any cosmological model we need to compare the theoretically calculated
distance modulus $\mu_{th}$ to the observable values (\ref{muobs}) from
Ref. \cite{Suzuki2012} using the relation 
redshift $z$:
\begin{equation}
\mu_{th}(z)= 5 \log_{10} \left(\frac{D_L (z)}{10\mbox{pc}}\right)
=5\log_{10}\frac{H_0D_L}c+\mu_0. 
 \label{mu}
\end{equation}
where $\mu_0=42{.}384-5\log_{10}h$, $D_L (z)$ is the luminosity distance \cite{Riess1, NP2005}
 \begin{equation}
 D_L(z)=\frac{c\,(1+z)}{H_0}S_k
 \bigg(H_0\int\limits_0^z\frac{d\tilde z}{H(\tilde
 z)}\bigg)  \label{DL} \end{equation}
 with $S_k$ having the form \cite{Sharov:2015ifa}
 
$$S_k(x)=\left\{\begin{array}{ll} \sinh\big(x\sqrt{\Omega_k}\big)\big/\sqrt{\Omega_k}, &\Omega_k>0,\\
 x, & \Omega_k=0,\\ \sin\big(x\sqrt{|\Omega_k|}\big)\big/\sqrt{|\Omega_k|}, &
 \Omega_k<0.
 \end{array}\right.$$
The quantity $H_0D_L/c$ in Eq.~(\ref{mu}) is the Hubble free luminosity distance 
and only the term $\mu_0$ \cite{NP2005} depend on the
Hubble constant $H_0$ and here we consider the spatially flat 
FLRW universe (i.e. $\Omega_k = 0$) as mentioned earlier. We use only the 
Union 2.1 compilation \cite{Suzuki2012} to constrain the particle creation models. 
The likelihood of the analysis follows $\mathcal{L} \propto e^{-\chi^2/2}$ and use the $\chi^2$-minimization technique, see for instance \cite{Sharov:2015ifa}. 

\begin{figure}
	\includegraphics[width=6.9cm, height=5cm]{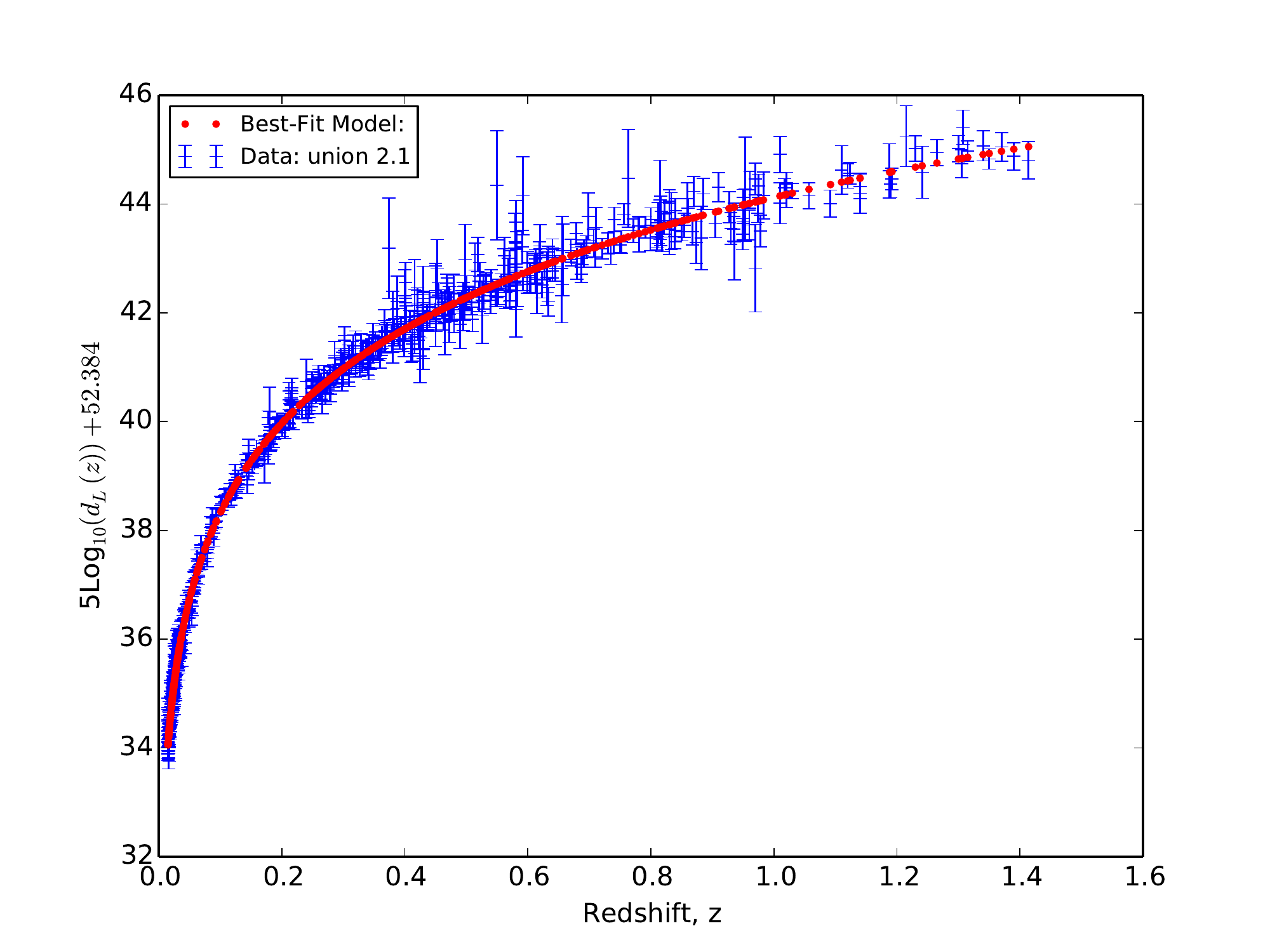}
	\includegraphics[width=6.9cm, height=5cm]{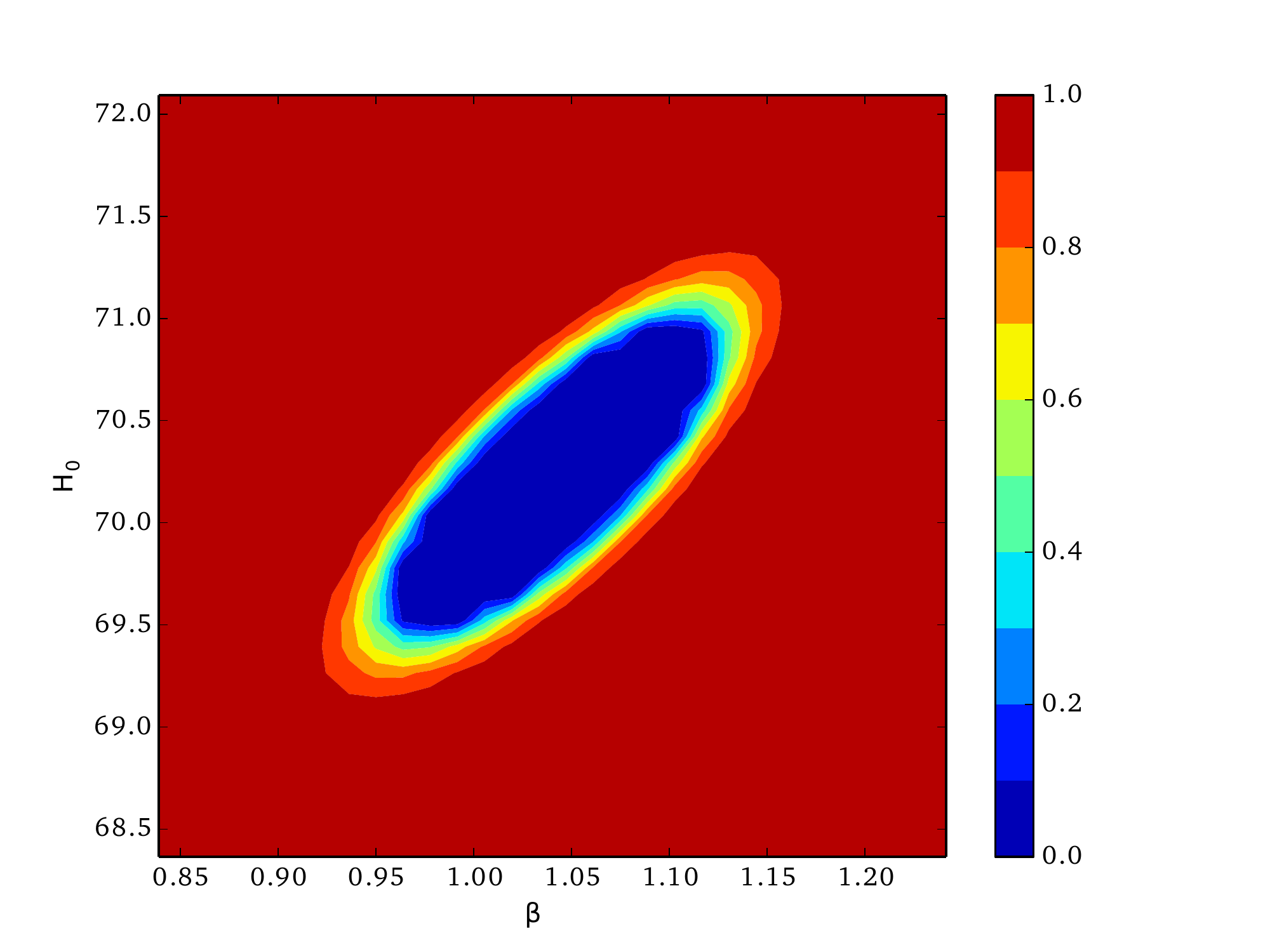}
	\caption{\label{best_fit_m2} Model I for best fit $H_0$ and $\beta$ along
		with the Union 2.1 data}
\end{figure}

\begin{figure}
	\includegraphics[width=6.9cm, height=5cm]{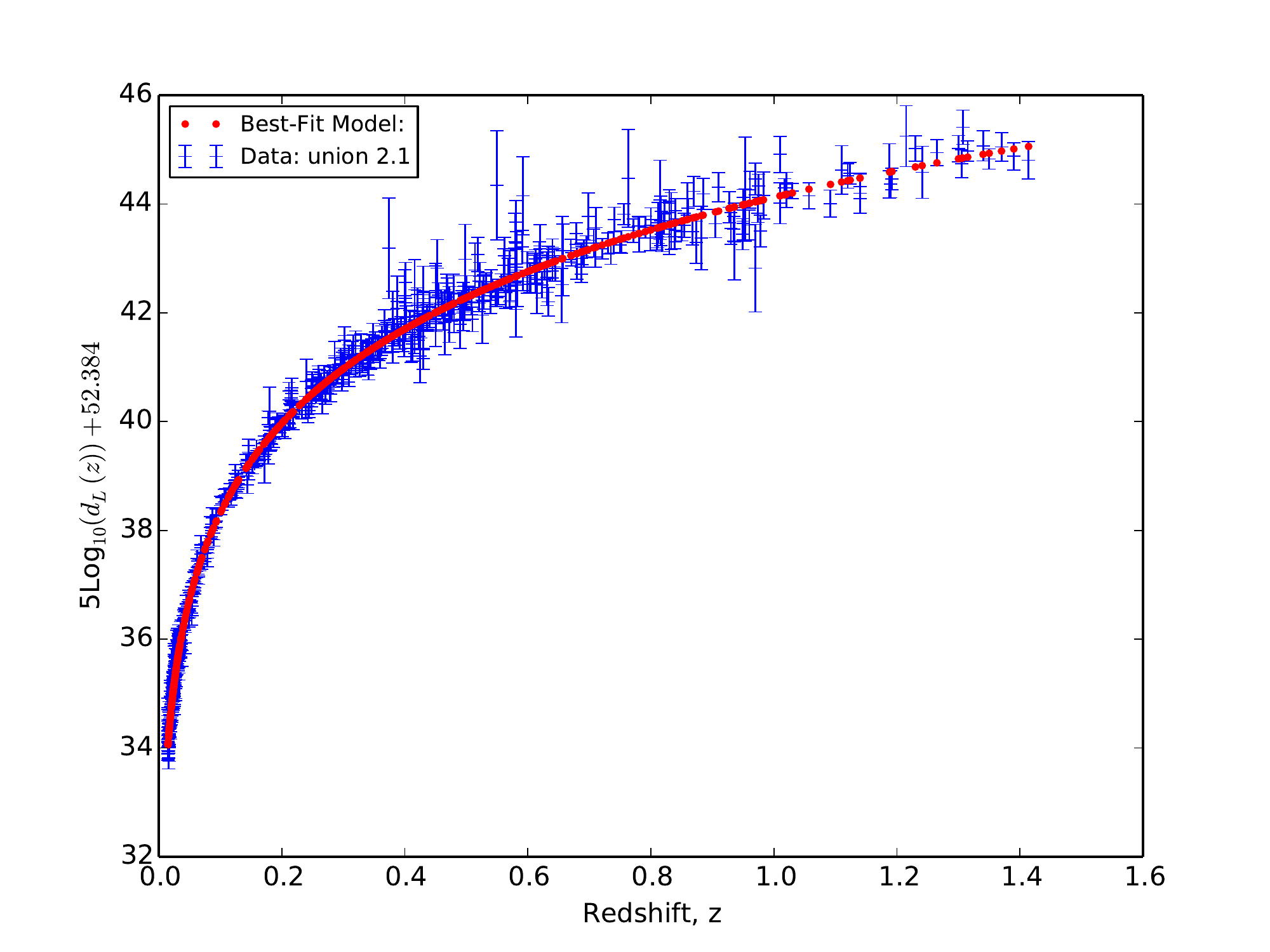}
	\includegraphics[width=6.9cm, height=5cm]{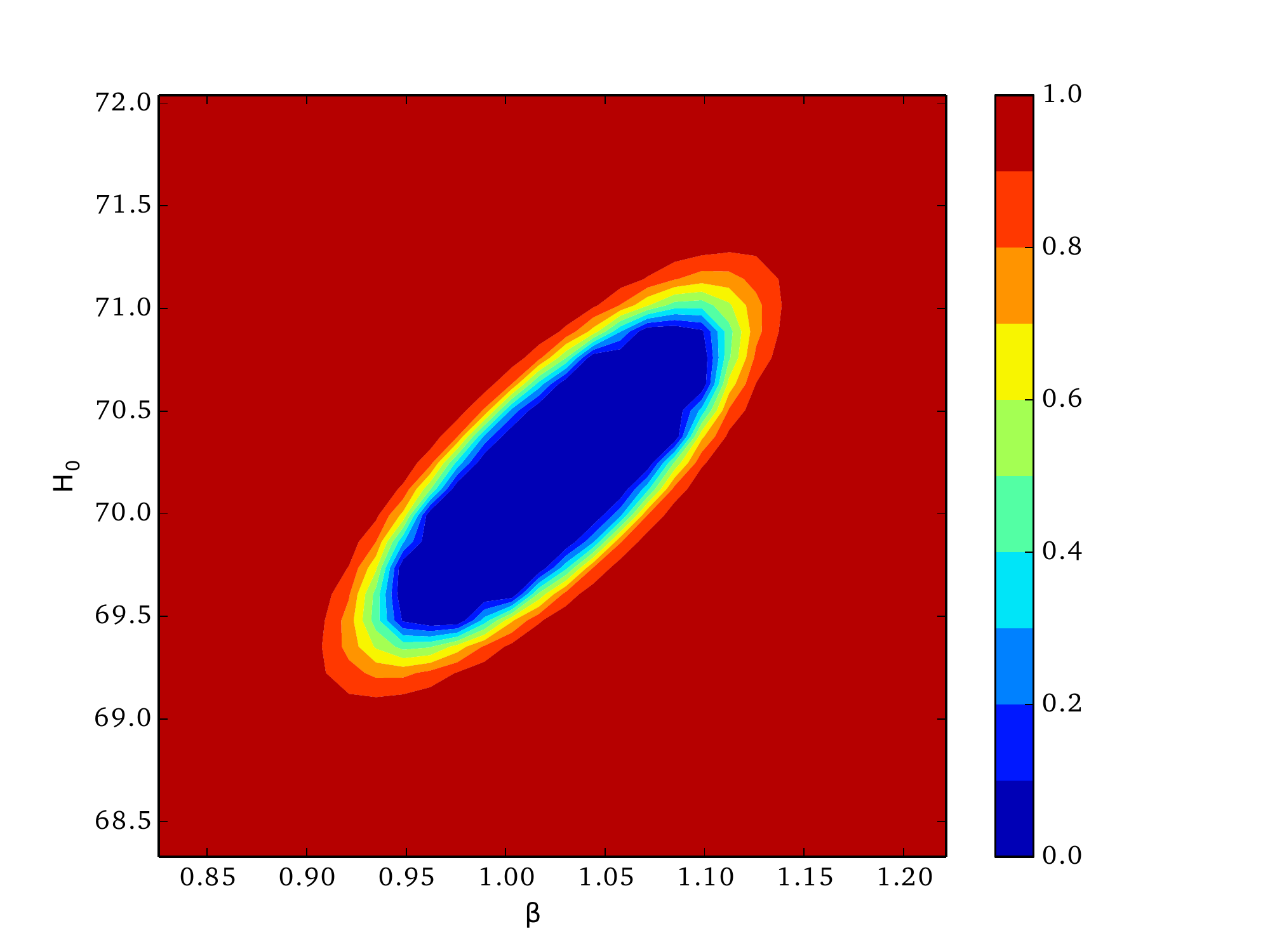}
	\caption{\label{best_fit_m3} Model II for best fit $H_0$ and $\beta$ along
		with the Union 2.1 data.}
\end{figure}

\begin{figure}
	\includegraphics[width=6.9cm, height=5cm]{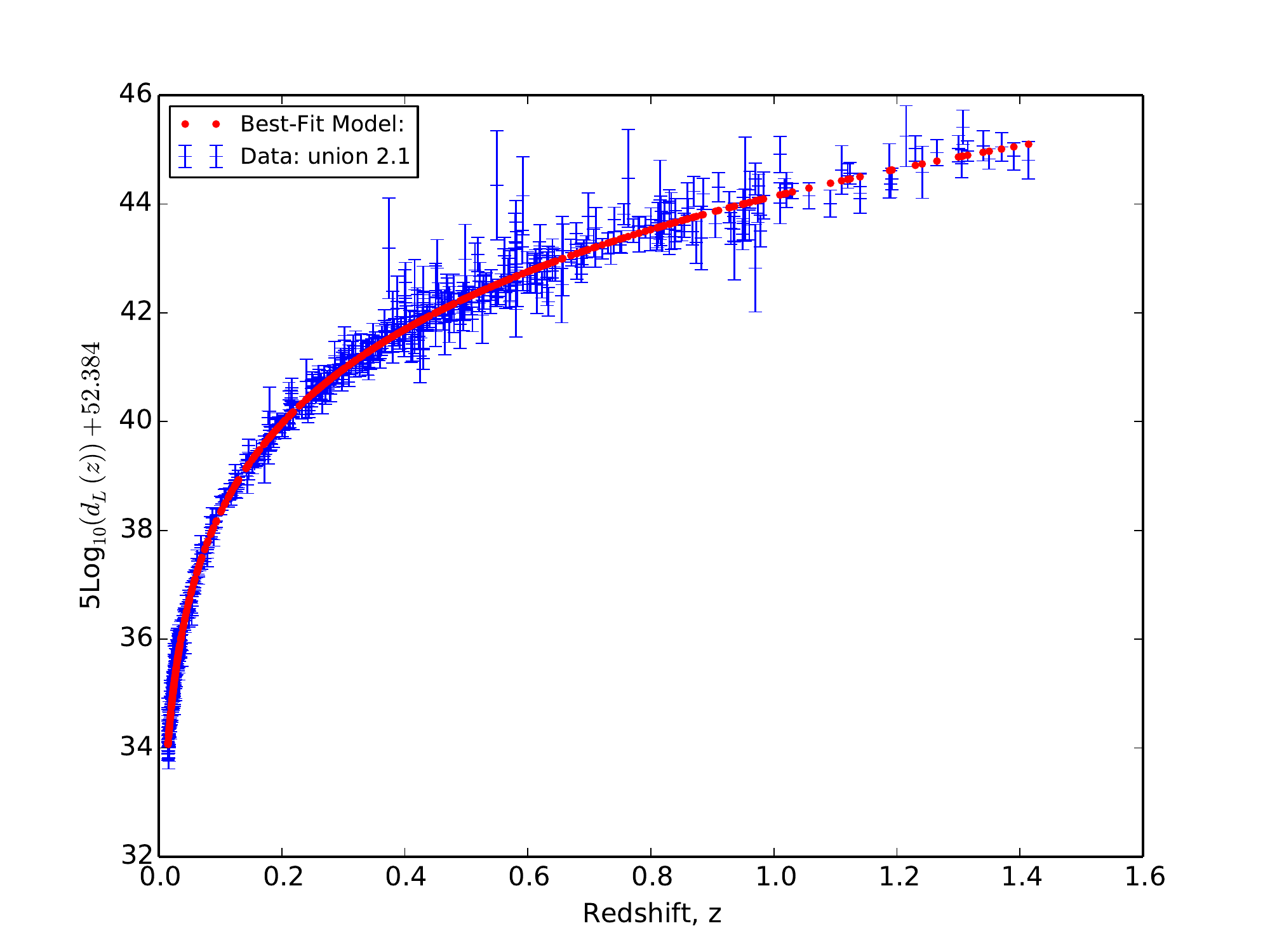}
	\includegraphics[width=6.9cm, height=5cm]{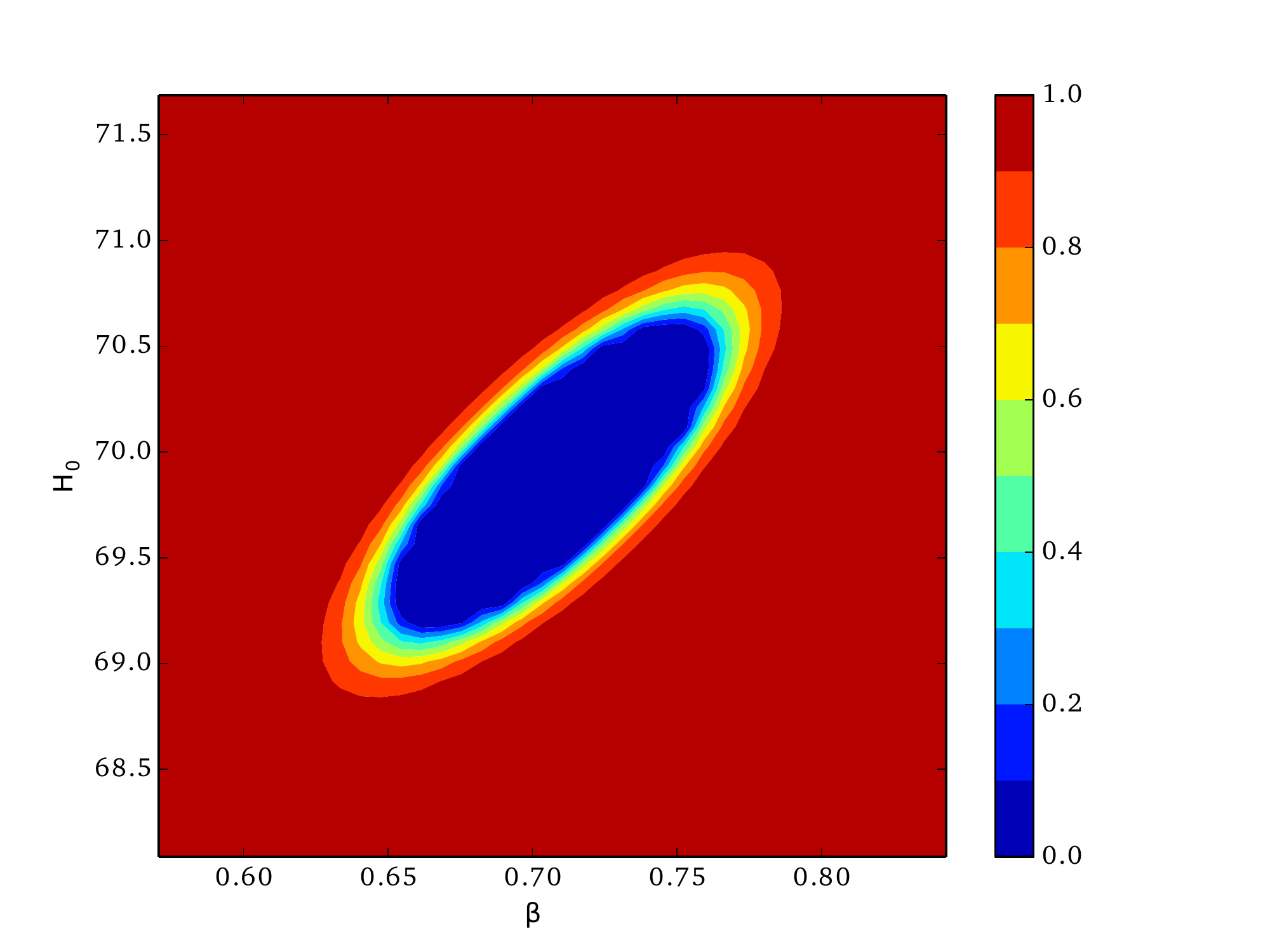}
	\caption{\label{best_fit_m5} Model III for best fit $H_0$ and $\beta$ along
		with the Union 2.1 data.}
\end{figure}

\begin{figure}
	\includegraphics[width=7.9cm, height=6cm]{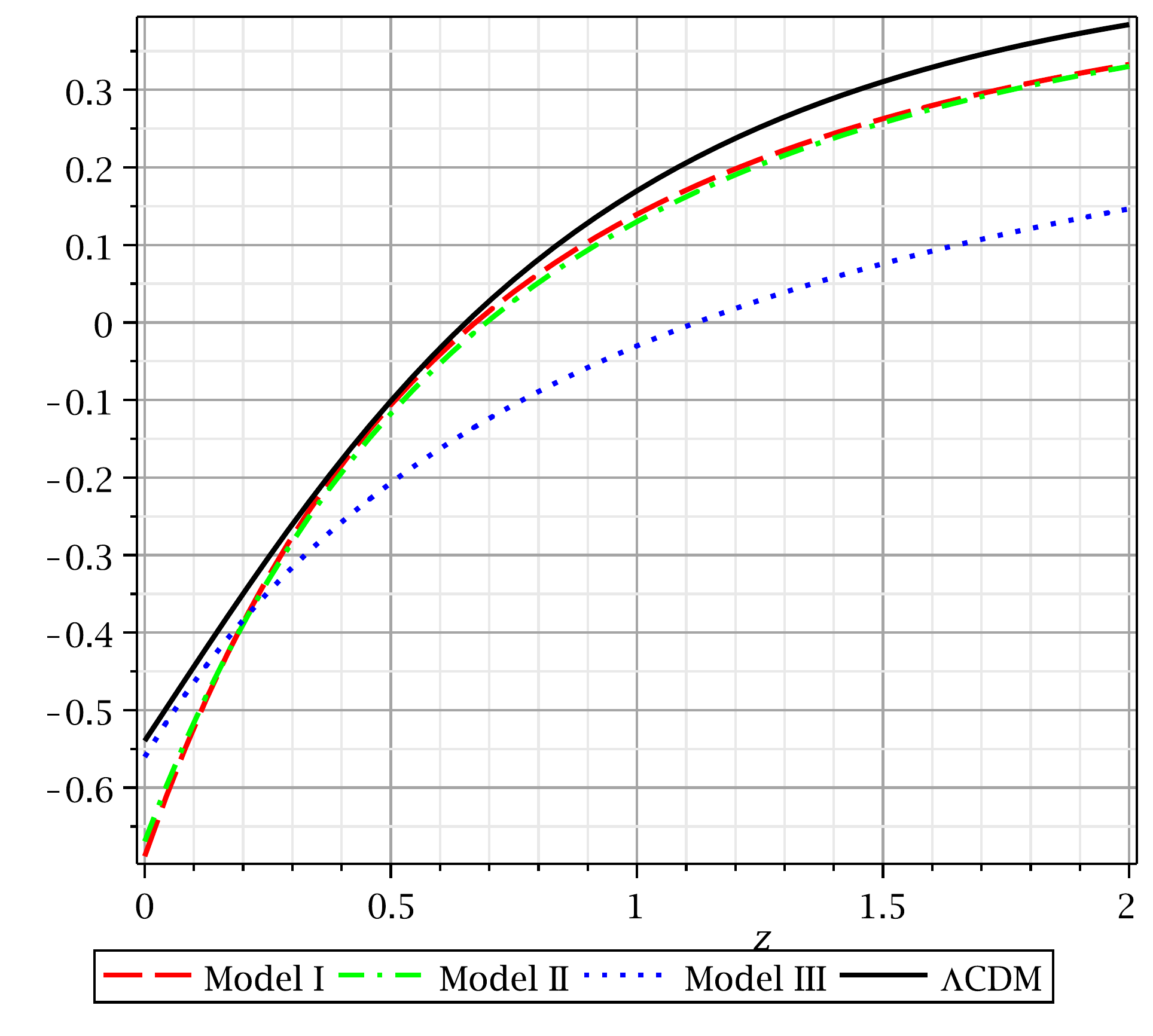}
	\caption{\label{figure-deceleration} using the best fit values of the model parameters, we display the transition of the deceleration parameters for three models in compared to the flat $\Lambda$CDM model.}
\end{figure}

We found that the parameter $\alpha$ does not make any significant contribution to the statistical analyses for models I and II because the function $\tanh (\alpha x)$ (where $\alpha, x \in \mathbb{R}$) is very slowly varying as mentioned earlier. So, we have fixed $\alpha = 1$ in models I and II and constrained models I and II as well as model III.   
Our analysis shows that for model I the best fit values of $H_0$ and $\beta$ are $H_0=69.9352\ {\rm km\ s^{-1} \ Mpc}^{-1}$, and $\beta=1.0403$. The constraint on $H_0$ is slightly high in compared to some latest observations \cite{Ade:2015xua}. For model II we have similar observations, that means here too we find that $H_0=69.8925\ {\rm km\ s^{-1} \ Mpc}^{-1}$ and $\beta=1.0235$. However, model III predicts a slight deviation from the previous models in $\beta=0.70693$, but its Hubble parameter value $H_0=69.88627\ {\rm km\ s^{-1} \ Mpc}^{-1}$ is comparable with the predictions from models II and III. In Figures \ref{best_fit_m2}, \ref{best_fit_m3}, \ref{best_fit_m5} we display the errors bars (left panel) with the best fit values in the plane ($H_0$, $\beta$) (right panel) with 1$\sigma$, 2$\sigma$ confidence levels.
Moreover, in order to understand the quantilative behaviour of the models, in figure \ref{figure-deceleration}, we show the evolution of the deceleration parameters for three particle creation models in compared to the spatially flat $\Lambda$CDM model. We find that models II and III almost matches with the evolution of the $\Lambda$CDM where the transition of the decelerating phase to the present accelerating phase occurs at around $z \sim 0.6$, whereas model III predicts the transition at around $z \sim 1.1$ which is significantly higher and hints a strong deviation from the other two models.


\section{Model independent tests}
\label{Sec-CP}
Model independent description is always fascinating for any kind of studies in nature, and when the discussion deals with our current accelerating universe, specifically with the dark energy models, it becomes  essential due to the large number of dark energy models in the literature. Two notable model independent geometrical, dimensionless parameters $\{r,~s\}$ are defined as \cite{Sahni}
\begin{eqnarray}
r&=& \frac{1}{aH^3}\frac{d^3 a}{dt^3},~~\mbox{and}~~s=\frac{r-1}{3\left(q-\frac{1}{2}\right)},\label{statefinder}
\end{eqnarray}
where $a$, $H$, $q$ have their usual meanings. The parameters are used to compare the goodness of several
dark energy models with the $\Lambda$CDM model, where for the  flat
$\Lambda$CDM, $\{r,s\}=\{1,0\}$. That means, throughout the evolution of the universe, $r (z_1)- r(z_2)= 0$, for any two arbitrary redshifts $z_1$, $z_2$. Later this model independent approach was further extended in Ref. \cite{Visser04} by considering the Taylor series expansion of the scale factor around the present time, and, which give rise some new model independent dimensionless geometrical parameters as follows:
\begin{eqnarray}
j= \frac{1}{aH^3} \frac{d^3 a}{dt^3},~s= \frac{1}{aH^4} \frac{d^4 a}{dt^4},~l= \frac{1}{aH^5}\frac{d^5 a}{dt^5},~\mbox{and}~m= \frac{1}{aH^6}\frac{d^6 a}{dt^6}.\label{CP}
\end{eqnarray}
The parameters in (\ref{CP}) together with $H$ and $q$ are called the cosmographic parameters\footnote{Here $r$ and $j$ are same; but the $s$ parameter defined in (\ref{CP}) is not same with one defined in (\ref{statefinder})}. The cosmographic parameters in Eq. (\ref{CP}) can be rewritten as
\begin{eqnarray}
j&=& - q+ (1+z) \frac{dq}{dz}+ 2 q(1+q),\label{j}\\
s&=& j- 3 j(1+q)- (1+z)\frac{dj}{dz},\label{s}\\
l&=& s- 4 s (1+q)- (1+z)\frac{ds}{dz},\label{l}\\
m&=& l-5 l (1+q)- (1+z)\frac{dl}{dz}.\label{m}
\end{eqnarray}
Therefore, from the above set of equations, we may argue that, as long as the Hubble parameter for any cosmological model is fourth order differentiable, the cosmography of that model will be valid.
\begin{table}
	\begin{tabular}{c c c c c c}
		\hline
		\hline
		Model \quad&\quad $q_0$ \quad&\quad $j_0$ \quad&\quad $s_0$  \quad&\quad$l_0$  \quad&\quad $m_0$ \\
		\hline
		\hline
		$\Lambda$CDM\quad&\quad $-0.53665$\quad &\quad 1.00000\quad&\quad$-0.39005$\quad&\quad $3.21486$\quad&\quad$-11.49597$\\
		Model I\quad&\quad $-0.56035$\quad&\quad$1.12798$\quad&\quad$2.01690$\quad&\quad$8.17418$\quad&\quad$23.78283$
		\\
		Model II\quad &\quad $-0.54115$\quad&\quad $0.93308$\quad &\quad $0.98352$\quad&\quad $3.37609$\quad&\quad$1.68084$\\
		Model III\quad& \quad$-0.56039$\quad& \quad$1.12809$\quad& \quad$2.01731$\quad&\quad $8.17558$\quad& \quad$23.79135$\\
		\hline
		\hline
	\end{tabular}
	\caption{The table shows the present values of deceleration ($q_0$) and all cosmographic parameters for the three models along with the flat $\Lambda$CDM model.}
	\label{table-CP}
\end{table}

\begin{figure}
	\includegraphics[width=6.9cm, height=5cm]{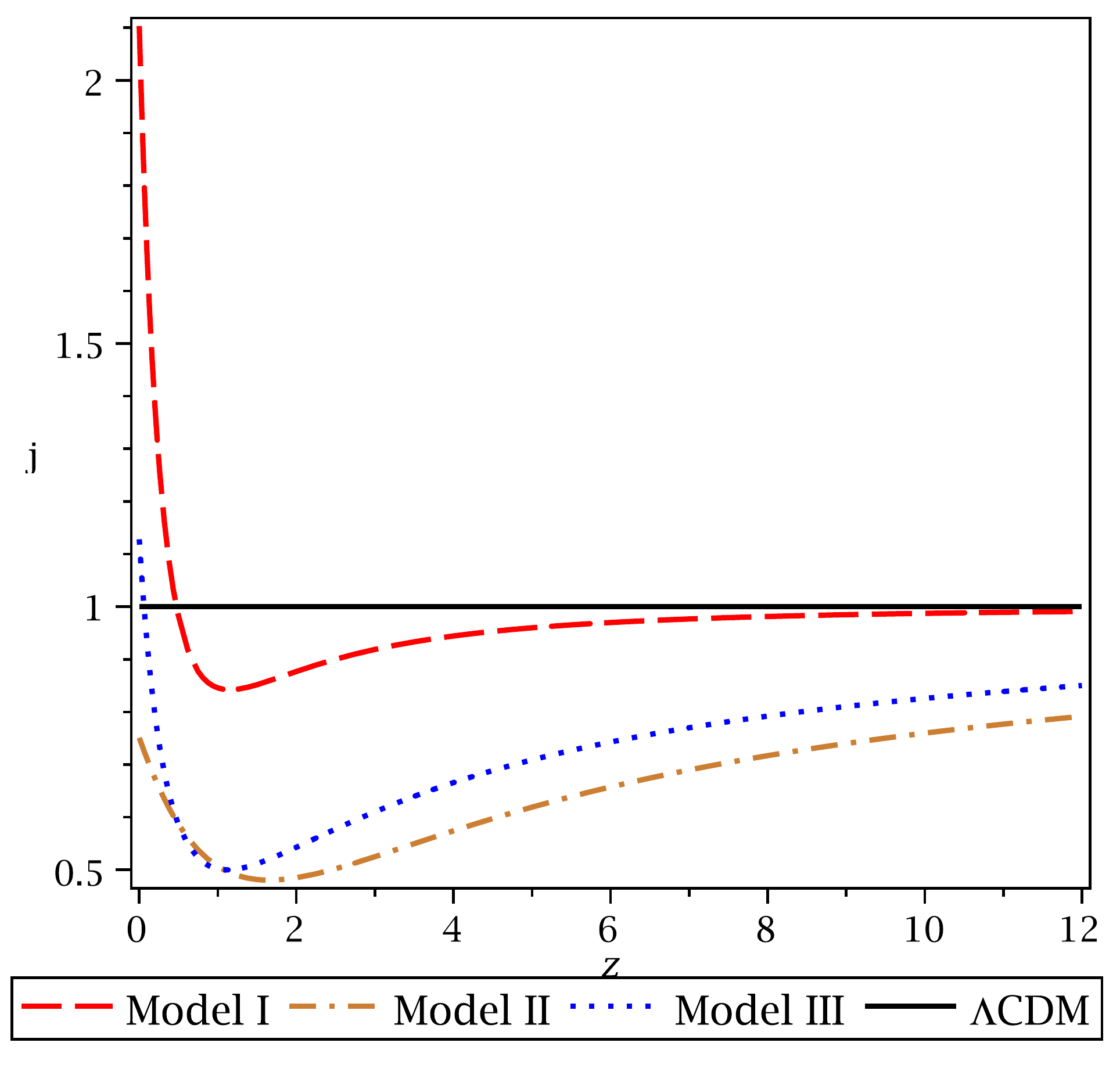}
	\includegraphics[width=6.9cm, height=5cm]{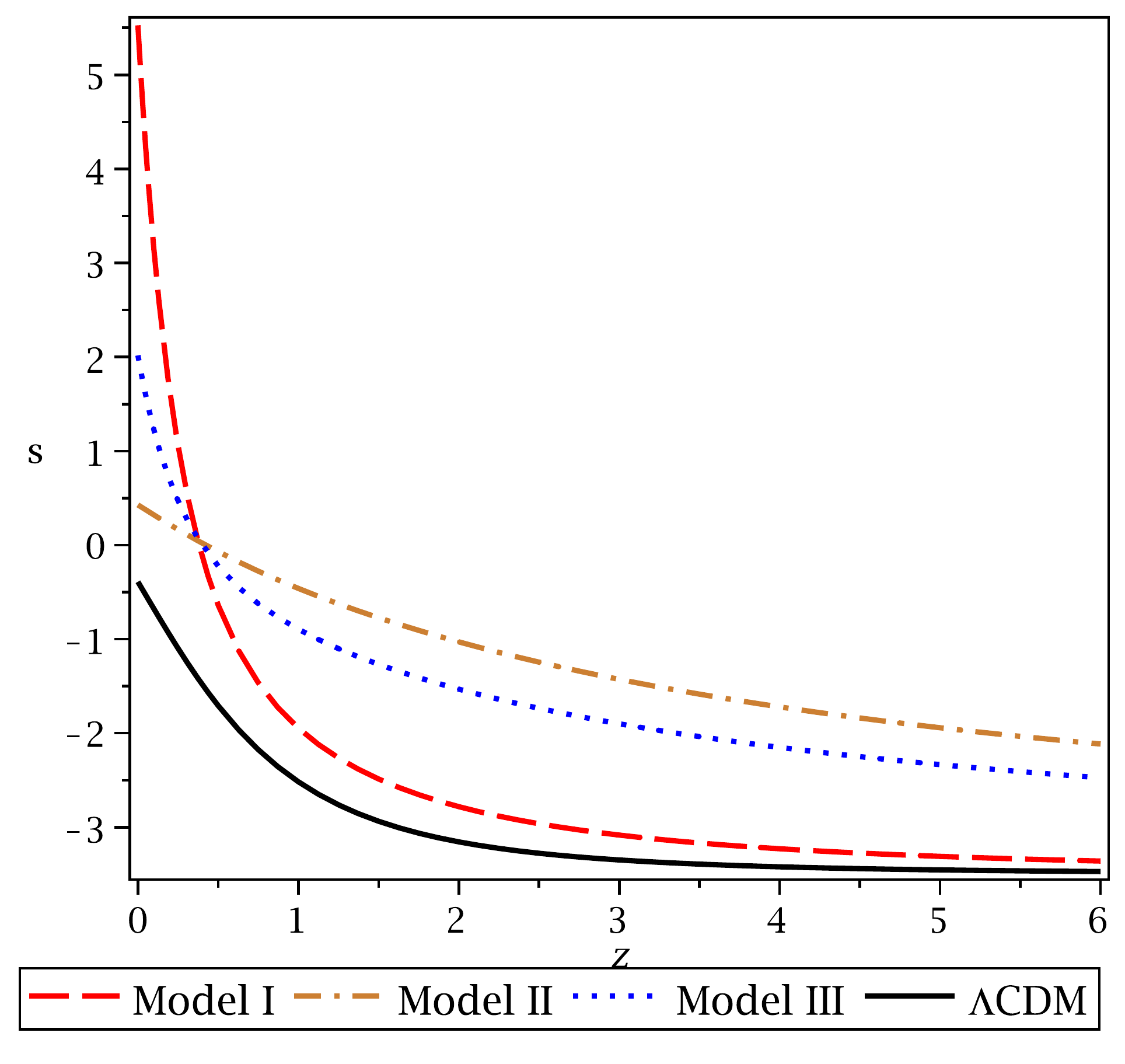}
	\includegraphics[width=6.9cm, height=5cm]{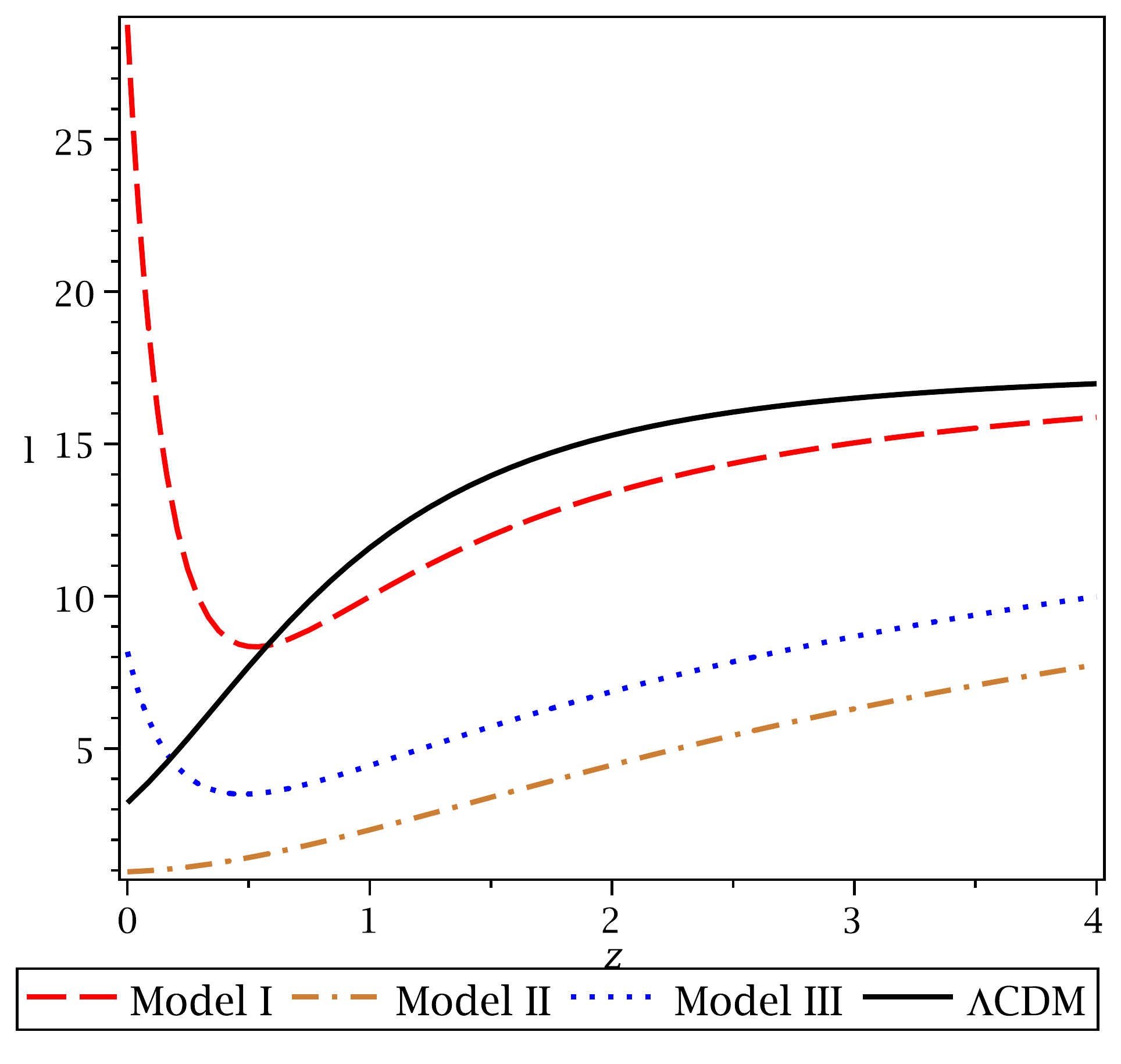}
	\includegraphics[width=6.9cm, height=5cm]{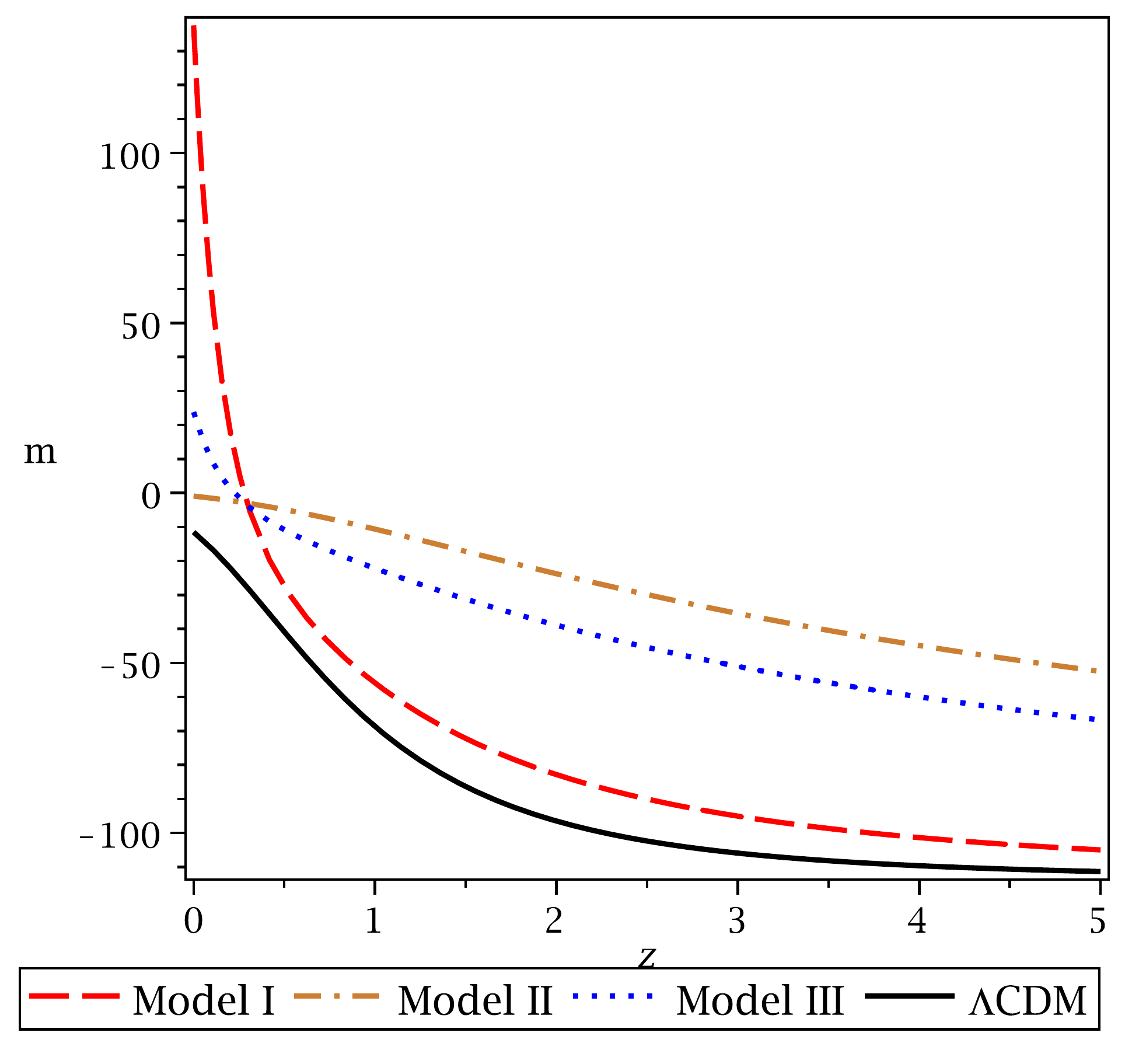}
	\caption{\label{figures-CP} The plots depict the evolution of the cosmographic parameters $j$, $s$, $l$, $m$ for three models in compared to the flat $\Lambda$CDM model}
\end{figure}

In Figure \ref{figures-CP}, we have shown the evolution of the cosmographic parameters for three particle creation models. In all the plots, we have kept the evolution of the corresponding cosmographic parameters for $\Lambda$CDM in order to compare the phenomenological models. From the figures (even the Table \ref{table-CP} also reflects the same nature), we find that, the parameters $j$, $l$ have similar evolution. On the other hand, the remaining parameters $s$, $m$ predict an equivalent evolution.

Through the statefinder parameters and its extension, the cosmography, we are able to  distinguish between several theoretically developed dark energy models from the $\Lambda$CDM model. In 2008, another model independent test $Om$ was introduced \cite{SSS2008} which can also differentiate the dark energy models from $\Lambda$CDM. The $Om$ function is defined as \cite{SSS2008}
\begin{eqnarray}
Om (z)&=& \frac{\tilde{h}^2 (z)-1}{(1+z)^3-1},~~~\mbox{where}~~\tilde{h}= \frac{H (z)}{H_0}.\label{om1}
\end{eqnarray}
Eq. (\ref{om1}) is very simple and elegant. It needs only the expansion rate to find the distinction of any dark energy model from the $\Lambda$CDM. It is easy to see that, for a spatially flat $\Lambda$CDM model, Eq.  (\ref{om1}) reduces to $Om (z)= \Omega_{m0}$ (where $\Omega_{m0}$ is the density parameter for the cold dark matter). That means, this function stays constant for flat $\Lambda$CDM model throughout the entire evolution of the universe. So, essentially, for a flat $\Lambda$CDM model, $Om(z_1)-Om(z_2)= 0$, for any two redshifts $z_1$, $z_2$. Therefore, compared to the statefinder parameter `$r$',
the behavior of $Om$ is almost same as both of them stay constant for flat $\Lambda$CDM, but, there is one notable property we should mention. The statefinder parameter $r$ needs triple derivative term with respect to the cosmic time, whereas as mentioned the construction of $Om$ needs only one time derivative with respect to the cosmic time. So, essentially, $Om$ is a more easier geometrical test in compared to the statefinders and cosmography.  However, one can see $Om$ as a two point function \cite{SSS2012} in the following way
\begin{eqnarray}
Om (z_i;z_j)&=& \frac{\tilde{h}^2(z_i)- \tilde{h}^2(z_j)}{(1+z_i)^3-(1+z_j)^3}.\label{om2}
\end{eqnarray}
Basically, Eq. (\ref{om2}) is nothing but  (\ref{om1}). If we simply put $z_j= 0$ in (\ref{om2}), we see $Om (z_i;0)= Om (z_i)$ which is Eq. (\ref{om1}). It is worthy to mention the elegant nature of Eq. (\ref{om2}). If we can know the values of the Hubble parameter at two or more redshifts, we can reconstruct $Om$ directly from the observations, and, consequently, this reconstruction will surely tells us whether the present universe is dominated by $\Lambda$CDM or not. However, Eq. (\ref{om2}) can be written in a sophisticated way by multiplying both sides of it by $h^2$ (where $h= 100^{-1} H_0$ km/sec/Mpc) as follows \cite{SSS2014}

\begin{eqnarray}
Om h^2 (z_i;z_j)&=& \frac{h^2(z_i)- h^2(z_j)}{(1+z_i)^3-(1+z_j)^3}~.\label{om3}
\end{eqnarray}
where $h(z)$ = $100^{-1}$ $H(z)$ km/sec/Mpc, and, $i \neq j$. For flat $\Lambda$CDM model,
we get

\begin{eqnarray}
Omh^2&=& \Omega_{m0} h^2~.\label{om4}
\end{eqnarray}
From the latest Planck mission \cite{Ade:2015xua}, $\Omega_{m0} h^2= 0.14170\pm 0.00097$ (TT,TE,EE$+$lowP$+$lensing$+$ext), thus, for the flat $\Lambda$CDM
model, one calculates that,

\begin{eqnarray}
Omh^2&=& 0.14170\pm 0.00097,\label{om5}
\end{eqnarray}
which is nothing but an indication to those dark energy candidates
which try to deviate from flat $\Lambda$CDM. Let us now consider three
redshifts $z_1= 0$, $z_2= 0.57$, $z_3= 2.34$, out of which the measurement of the Hubble parameters at $0.57, ~2.34$ are statistically independent. Thus, they will be very much helpful in order to calculate $Om h^2(z_i;z_j)$. In Table \ref{table-om} we summarize the values of $Omh^2 (z_i;z_j)$ for the particle creation models which makes a comparison between the models as welll as with $\Lambda$CDM

\begin{table}
	\begin{tabular}{c c c c c}
		\hline
		\hline
		$Om h^2 (z_i;z_j)$\quad&\quad $\Lambda$CDM\quad&\quad Model I \quad&\quad Model II \quad&\quad Model III\\
		\hline
		\hline
		$Omh^2(z_1; z_2)$\quad&\quad $0.14170\pm 0.00097$\quad&\quad 0.05729\quad &\quad0.05696\quad&\quad0.05762\\
		$Omh^2(z_1;z_3)$\quad&\quad $0.14170\pm 0.00097$\quad&\quad 0.02892 \quad&\quad0.02831\quad&\quad 0.02906\\
		$Omh^2(z_2;z_3)$\quad&\quad $0.14170\pm 0.00097$\quad&\quad 0.01576\quad &\quad 0.01521\quad &\quad 0.01582\\
		\hline
		\hline
	\end{tabular}
	\caption{The table shows the $Omh^2$ values for the models I, II, and III in compared to the flat $\Lambda$CDM model.}
	\label{table-om}
\end{table}

\section{Thermodynamic restrictions}
\label{sec:thermodynamics}
The viability of a cosmological model is tested from its thermodynamical behaviour. A profound relation between gravity and thermodynamics has been already found \cite{bekenstein1, bekenstein2, hawking1, hawking2, jacobson, paddy}, so it is very natural to investigate the thermodynamical bounds on any specific cosmological model. A concise description on the thermodynamic propoerties of the modified gravity theories can be found in \cite{Bamba:2016aoo}. 
We devote this section in order to check the conditions of thermodynamic equilibrium for the present particle production models. Since from the physical point of view, macroscopic systems tend toward the thermodynamic equilibrium and
the entropy ($S$) of an macroscopic system is never decreasing. So, from mathematical point of view, if $S$ is the total entropy of the macroscopic system, then we must have $\dot{S} \geq 0$ (Entropy never decreasing), and $\ddot{S} < 0$ (Equilibrium condition) for $t \longrightarrow \infty$. In other words, the entropy should be concave in a small neighborhood of the equilibrium point. Now, the total entropy ($S$) is contributed from the entropy of the apparent horizon ($S_h$) and the entropy of the fluid ($S_w$) with the equation of state $p= w \rho$, that means, essentially, we have to consider the behavior of $S= S_h+ S_w$. Now, the entropy of the apparent horizon is given by $S_h = k_{B} \mathcal{A}/4\, l_{pl}^2$, where $k_{B}$ is the Boltzmann's constant, $\mathcal{A}= 4 \pi r_h^2$, is the horizon area in which $r_h= \left(H^2+ k/a^2 \right)^{-1/2}$ is the horizon radius \cite{Bak} which finally becomes  $r_h= 1/H$.  Now, differentiating $S_h$ with respect to the cosmic time and remembering the fact
that we are considering flat FLRW universe,  one gets

\begin{align}\label{th1}
\dot{S}_h & = -\, \left(\frac{2 \pi k_B}{l_{pl}^2\, H^3}\right)\, \dot{H}  = \frac{3 \pi k_B}{l_{pl}^2 H}\, (1+w) \left(1-\frac{\Gamma}{3H} \right),
\end{align}
which shows that for $\dot{S}_h \geq 0$, we should have  $\Gamma/ 3H \leq 1$. Now, we recall the Gibb's equation for the fluid which is a relation between the thermodynamic quantities associated with the fluid as $T dS_{w}= d (\rho\, V) + p dV$, where by $S_w$ we mean the entropy of the fluid, $V= 4\pi r_h^3/ 3$ is the volume of the region surrounded by the radius $r_h$, and $T$ is the fluid temperature. Now, encountering the cosmic time in the Gibb's equation one may come at

\begin{align}\label{fluid-entropy}
T\dot{S}_{w} &=6\,(1+w)\,\pi\,\left(1-\frac{\Gamma}{3H}\right)(1+ 3w).
\end{align}
Since for $ w> 0$, we must have that $\dot{S}_w \geq 0$. Finally, one has that $\dot{S}_h+ \dot{S}_w \geq 0$ for $\Gamma/ 3H \leq 1$. So, the entropy of the horizon plus the fluid is an increasing function of the cosmic time. Now, we proceed for the equilibrium condition, and we need to define the temperature of the fluid first. If one assumes that the temperature of the fluid becomes equal to that of the temperature of the horizon defined as $T_h= 1/2\pi r_h$ \cite{temperature} then one may find the second derivatives of $S_w$. However,
differentiating $\dot{S}_h$ again with respect to the cosmic time, we find

\begin{align}\label{1}
\ddot{S}_h & = \left(\frac{3 \pi k_B (1+w)}{l_{pl}^2}\right)\, \left[ \frac{3}{2} (1+w) \left( 1-\frac{\Gamma}{3H} \right)\, \left( 1-\frac{2\Gamma}{3H} \right) -\frac{1}{3H}\, \frac{\dot{\Gamma}}{H^2} \right].
\end{align}
Similarly, if one differentiates eq. (\ref{fluid-entropy}) with respect to the cosmic time, then one gets

\begin{align}\label{2}
\ddot{S}_w & = 12 \pi^2 (1+w) (1+3w) \left[ \frac{3}{2} (1+w) \left( 1-\frac{\Gamma}{3H} \right)\, \left( 1-\frac{2\Gamma}{3H} \right) -\frac{1}{3H}\, \frac{\dot{\Gamma}}{H^2} \right].
\end{align}
Now, adding equations (\ref{1}) and (\ref{2}), one has

\begin{align}\label{sum}
\ddot{S}_h + \ddot{S}_w & = \left( \frac{3 \pi k_B (1+w)}{l_{pl}^2}+ 12 \pi^2 (1+w) (1+3w) \right)\,\,\left[ \frac{3}{2} (1+w) \left( 1-\frac{\Gamma}{3H} \right)\, \left( 1-\frac{2\Gamma}{3H} \right) -\frac{1}{3H}\, \frac{\dot{\Gamma}}{H^2} \right].
\end{align}
Now, the key role in determining the sign of $S= \ddot{S}_h + \ddot{S}_w$ is played by the second bracketed term in equation (\ref{sum}). If one proceeds further, then after some simple steps, one may conclude that $\ddot{S}< 0$, provided the rate $\Gamma$ satisfies the following simple relation

\begin{align}\label{ineq1}
\frac{d\Gamma}{dH} + 3H \left( 1-\frac{2\Gamma}{3H} \right) & < 0\,,
\end{align}
which can be considered as a very general relation for all particle production models. From here, since we have introduced the rate of change of production rate $\Gamma^\prime \equiv \frac{d\Gamma}{dH}$, so one may encounter with two possible cases as follows: It may happen that either $\Gamma^\prime> 0$, or $\Gamma^\prime< 0$. However, one may notice that $\Gamma/H$ plays an important role since we have already noticed that this term deviates the evolution equations from their standard laws, and also for $\Gamma/3H \ll 1$, one gets back the standard evolution, so we express the inequality in eq. (\ref{ineq1}) by the following inequality

\begin{align}\label{ineq2}
\frac{d}{dt} \left(\frac{\Gamma}{H}\right) & > \frac{3}{2}\,(1+w)\, \left(1-\frac{\Gamma}{3H}\right)\, \left[\frac{\Gamma}{H}+ 3H \left(1- \frac{2\Gamma}{3H}\right) \right].
\end{align}
The prescription is very general in the sense that it provides a thermodynamic constraints over any particle production model. Since we have considered three different particle production models given in (\ref{M1}), (\ref{M2}), and (\ref{M3}), we find that the generalized second law of thermodynamics is always valid for $\Gamma/ 3H < 1$, and the models will be thermodynamically stable if one of the inequalities in equations (\ref{ineq1}), (\ref{ineq2}) holds good.\newline

Now, we shall investigate the same thermodynamical properties when the temperature of the fluid is governed by the following law \cite{Zimdahl2, LB14}

\begin{align}\label{temp2}
	\frac{\dot{T}}{T} & = 3H \left( \frac{\Gamma}{3H} - 1\right)\frac{\partial p}{\partial \rho}.
\end{align}
Since for our model $\frac{\partial p}{\partial \rho}= w$, so using the evolution equation (\ref{eqn10.1}) one can solve the above equation (\ref{temp2}) as

\begin{align}\label{temp2a}
	T &=T_0\left( \frac{H}{H_0} \right)^{2\,w/(1+ w)},
\end{align}
where $T_0$, $H_0$ are respectively  the present values of the temperature and the Hubble parameter. One may note that the new temperature the in (\ref{temp2a}) could effect on the equilibrium condition not on the generalized second law of thermodynamics, since the temperature is positive as observed from equation (\ref{temp2a}), so from equations (\ref{th1}), (\ref{fluid-entropy}), it is clear that both $\dot{S}_h$ and $\dot{S}_w$ are non-negative for $\Gamma/ 3H \leq 1$, thus their addition too, hence the generalized second law of thermodynamics is valid. Now, we need to calculate only $\ddot{S}_w$ since $\ddot{S}_h$ will remain same as it is in equation (\ref{1}). So, using the temperature  (\ref{temp2a}) into equation (\ref{fluid-entropy}), one may find that

\begin{align}\label{temp2b}
\ddot{S}_w & = -\frac{6\pi}{T_0} \, (1+w) (1+3w)\left[\frac{3H}{2}(1+3w) \left(1-\frac{\Gamma}{3H} \right) \left( \frac{\Gamma}{3H} -\frac{2w}{1+w} \right) + \frac{\dot{\Gamma}}{3H}\right].
\end{align}
Although it  seems difficult to arrive at some conclusion from the sum  $\ddot{S}_h + \ddot{S}_w$ on the thermodynamic equilibrium, but we find that under the following conditions the model could reach the thermodynamic equilibrium: (A) If the inequality (\ref{ineq1}) holds good together with  $\dot{\Gamma} < 0$, and $\Gamma/3H < 2w/(1+w)$ or (B) The inequality (\ref{ineq1}) should also again hold together with 
\begin{eqnarray}\label{ineq3}
\frac{d\Gamma}{dH}  < \left(\frac{1+3w}{1+w}\right) \left(\frac{\Gamma}{3H}- \frac{2w}{1+w}\right).
\end{eqnarray}
Thus, we find that under the conditions described above, the generalized second law of thermodynamics holds good for $\Gamma/3H \leq 1$ irrespective of the temperature of the fluid, but on the other hand, the thermodynamic equilibrium is reached under the specified conditions which significantly depends on the temperature as we have considered in this section. So, the generalized second law of thermodynamics holds for models I, II, III provided $\Gamma/3H\leq1.$


\section{Summary}
\label{discussion}

The theory of adiabatic particle creation by the time-varying gravitational field
has been intensively investigated to explain the late-time accelerated expansion of the universe.  In addition, such mechanism can also take into account of the early inflationary evolution as well as the intermediate phases of the cosmic evolution. The most remarkable feature in such theory is that, the description of the current acceleration of the universe does not need any dark energy fluid or modified gravity theories. However, this theory is also model dependent as dark energy or modified gravity theories. Thus, considering this fact, in this work we propose some new particle creation models which generalize the existing models in the literature and 
constrain them using the Union 2.1 compilation from Supernovae Type Ia data in order to 
extract the information from the models at low-redshifts. Our analysis shows that 
at low-redshifts the models are close to the standard $\Lambda$CDM model. We note that in all models,
a slightly higher values of the present Hubble 
parameter value are favored and the 
parameter $\beta$ is constrained to be close to unity for model I and II while 
$ 0< \beta < 1$ for model III.  Further, using the model independent diagnosis, 
we show that models I  and II are very close to each other and 
as well as with the $\Lambda$CDM model. We also study the thermodynamical
laws  for the particle creation models where primarily we establish the 
general conditions for any particle creation model 
that ensure the validity of the generalized second law of thermodynamics 
as well as the conditions for themrmodynamic equilibrium.

\section*{Acknowledgments}

The authors thank the referee for some important comments to improve the mansucript. The authors also thank Prof. J de Haro, Dr. A. Paliathanasis and Prof. J. A. S. Lima for helpful discussions.

\end{document}